\documentclass{aa}

\usepackage{lineno,hyperref}

\usepackage{natbib}
\usepackage{subfig}

\usepackage{amsmath}    
\usepackage[dvipsnames]{xcolor} 
\usepackage[T1]{fontenc}
\usepackage{ae,aecompl}
\usepackage[english]{babel}

\usepackage{hyperref}

\usepackage{url}
\usepackage{soul}

\usepackage{graphicx}   
\usepackage{amsmath}    
\usepackage{amssymb}    
\usepackage{adjustbox}
\usepackage{threeparttable} 
\usepackage{multirow} 
\usepackage{makecell} 
\usepackage[normalem]{ulem}
\usepackage{colortbl} 

\newcommand{\be}{\begin{equation}}
\newcommand{\ee}{\end{equation}}
\newcommand{\bea}{\begin{eqnarray}}
\newcommand{\eea}{\end{eqnarray}}

\DeclareMathAlphabet\mathbfcal{OMS}{cmsy}{b}{n}
\definecolor{Gray}{gray}{0.85}
\newcolumntype{a}{>{\columncolor{Gray}}c}

\usepackage{color}
\usepackage{todonotes}

\hypersetup{colorlinks=true, citecolor=blue, linkcolor=blue}

\modulolinenumbers[5]
\usepackage[section]{placeins}

\begin{document}

\title{Unveiling hidden companions in post-common-envelope binaries: A robust strategy and uncertainty exploration}

   \author{Cristian A. Giuppone\inst{1,2}\fnmsep\thanks{E-mail: \href{mailto:cristian.giuppone@unc.edu.ar}{cristian.giuppone@unc.edu.ar}},
        Luciana V. Gramajo\inst{2,3},
        Emmanuel Gianuzzi\inst{1,2},
        Matías N. Ramos \inst{1,2},
        Nicolás Cuello\inst{4},
        Tobias C. Hinse\inst{5}
        }

   \institute{
   Instituto de Astronom\'ia Te\'orica y Experimental (IATE), CONICET-UNC, C\'ordoba, Argentina,
   \and
   Observatorio Astron\'omico de C\'ordoba (OAC), UNC, C\'ordoba, Argentina,
   \and
   Consejo Nacional de Investigaciones Cient\'ificas y T\'ecnicas (CONICET), Godoy Cruz 2290, Ciudad Aut\'onoma de Buenos Aires, Argentina,
   \and
   Univ. Grenoble Alpes, CNRS, IPAG, 38000 Grenoble, France,
   \and
    University of Southern Denmark, Department of Physics, Chemistry and Pharmacy, SDU-Galaxy, Campusvej 55, DK-5230 Odense M, Denmark
   }

\date{}

\abstract
{Some post-common-envelope binaries (PCEBs) are binary stars with short periods that exhibit significant period variations over long observational time spans. These eclipse timing variations (ETVs) are most likely to be accounted for by the presence of an unseen massive companion, potentially of planetary or substellar nature, and the light-travel time (LTT) effect. The existence of such companions challenges our current understanding of planetary formation and stellar evolution.} 
{In this study, our main objective is to describe the diversity of compatible nontransit companions around PCEBs and explore the robustness of the solutions by employing tools for uncertainty estimation. We select the controversial data of the QS Vir binary star, which  previous studies have suggested hosts a planet.}
{We employ a minimizing strategy, using genetic algorithms to explore the global parameter space followed by refinement of the solution using the simplex method. We evaluate errors through the classical Markov chain Monte Carlo (MCMC) approach and discuss the error range for parameters, considering the $1\sigma$ values obtained from the minimization.}
{Our results highlight the strong dependence of ETV models for close binaries on the dataset used, which leads to relatively loose constraints on the parameters of the unseen companion. We find that the shape of the $O-C$ curve is influenced by the dataset employed. We propose an alternative method to evaluate errors on the orbital fits based on a grid search surrounding the best-fit values, obtaining a wider range of plausible solutions that are compatible with goodness-of-fit statistics. We also analyze how the parameter solutions are affected by the choice of the dataset, and find that this system continuously changes the compatible solutions as new data are obtained from eclipses.}
{The best-fit parameters for QS Vir correspond to a low-mass stellar companion (57.71 $M_{\rm jup}$ ranging from $\sim$40 to $\sim$64 $M_{\rm jup}$) on an eccentric orbit ($e=0.91^{+0.07}_{-0.17}$) with a variety of potential periods ($P = 16.69 ^{+0.47}_{-0.42}$ yr.). Most solutions within $1\sigma$ exhibit regular orbits, despite their high eccentricity. Additional observations are required to accurately determine the period and other parameters of the unseen companion. In this context, we propose that a fourth body should not be modeled to fit the data, unless new observations considerably modify the computed orbital parameters. This methodology can be applied to other evolved binary stars suspected of hosting companions.}
{}



\titlerunning{Unveiling hidden companions in Post-Common Envelope Binaries}

\authorrunning{Giuppone et al.}

\maketitle


\section{Introduction}\label{sec1}

The formation and evolution of main sequence (MS) binary star systems imply substantial changes in the shape and size of the orbit, and in the physical properties of the stellar components. However, the formation of stellar multiples is complex \citep{vandenberk2007}. Following a recent review by \citet{tokovinen2021}, three main ingredients are necessary for the formation of stellar systems: fragmentation, accretion, and dynamical processes. This involves the initial collapse of a {molecular} cloud of gas, and dust. The frequent gravitational interactions {and the subsequent} dynamical capture of nearby stars naturally leads to a high fraction of stellar multiplicity \citep{Larson_1972,vandenberk2007,Bate_2018}. Surprisingly, the formation of isolated binary systems is the exception and the majority of stars are found in hierarchical multiple stellar systems. In this context, post-common-envelope binaries (PCEBs) constitute a particular class of binaries with periods of the order of hours. Remarkably, some of them are believed to harbor one or more massive companion. The literature lists several systems where authors have claimed the discovery of one or two high-mass circumbinary companions of planetary nature with a semi-major axis of greater than $1\,\mathrm{au}$ \citep[see e.g.,][and references therein]{Marsh+2018}. 

To date, over 5000 confirmed planets have been discovered\footnote{\url{https://exoplanetarchive.ipac.caltech.edu/}}, with the majority found in orbit around a single host star. Approximately $5\%$ of these planets are known to orbit binary star systems \citep[\footnote{\url{https://www.openexoplanetcatalogue.com/systems/?filters=multistar}}][]{Schwarz+2016, Bonavita+2020}. In the case of PCEBs, the presence of low-mass stellar objects is indirectly detected through the effect of eclipse timing variation \citep[][ETV]{Woltjer+1922} caused by the light-travel time \citep[][LTT]{Irwin+1952} effect, also known as the R{\o}mer effect. Several other mechanisms that cause ETVs are known and we refer to \citet{southworth2019} for a review. According to \citet{Zorotovic+2013}, nearly 90$\%$ of PCEBs exhibit an ETV signal amplitude compatible with a substellar high-mass interpretation, although alternative period variation mechanisms have been suggested. One such mechanism that may account for the observed timing variation is the Applegate mechanism \citep{Applegate+1992, Volschow+2018}.

Additional physical mechanisms, such as mass transfer, apsidal precession, chromospheric spots, solar-like cycles in one of the stars, or angular momentum-loss due to magnetic breaking, may cause period variations \citep[see e.g.,][]{Lanza+2020, Pulley2022, lee2009}. In terms of magnitude, the largest contribution to period variation is from a bound massive companion. The effect of smaller period-variation contributions are usually accounted for by considering a quadratic term that describes secular changes in the binary eclipse period \citep{southworth2019}.

\begin{figure}
\begin{center}
\includegraphics[width=6 cm]{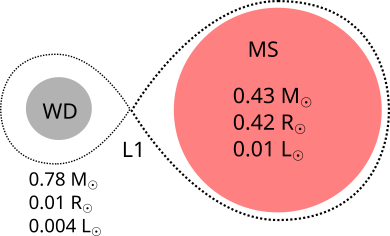}\\
\includegraphics[width=6 cm]{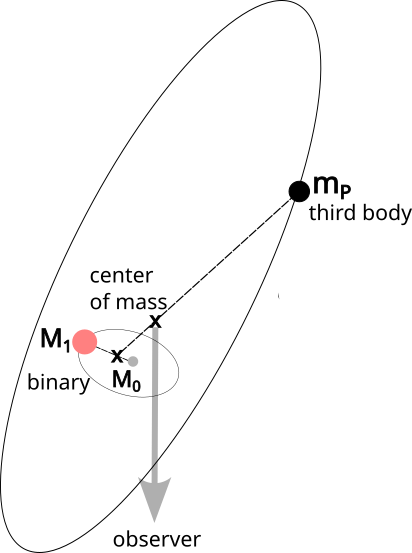}\\
\caption{Schematic representation of QS Vir with its physical parameters. Top: Physical parameters, Lagrangian point $L1$, and equipotential surface in QS Vir. The parameter values of the binary system were obtained by \citet{ODonoghue+2003} and we calculated the luminosity using temperature and radius. Bottom: Schematic configuration in the case where a third body modifies the binary barycentric position with respect to the observer. The movement of the binary can be modeled to infer the mass and the orbital parameters of the third body. 
}
\label{Fig-1}
\end{center}
\end{figure}

In the present work, we revisit the QS Vir system \citep[]{Stobie+1997} and present a time-series analysis for the determination of orbital fits based on the LTT model. The QS Vir binary system \citep[$V \simeq 11.8$,][]{ODonoghue+2003, Zacharias+2013} is composed of a white dwarf (WD) and a low-mass M-dwarf that nearly fills its Roche lobe (see Figure \ref{Fig-1}) with an orbital period of around 3.5~h \citep{ODonoghue+2003}. This binary system is classified according to its evolutionary phase as a PCEB, although it can also be associated with the precataclysmic variable (CV) classification \citep{Parsons2011a,Parsons2016}. This latter association is based on the fact that it is on its way to becoming a semi-detached cataclysmic variable through a greater loss of angular momentum and a reduction of the orbit due to gravitational waves or magnetic braking\citep{Paczynski1967, ODonoghue+2003}.

A schematic representation of QS Vir is shown in Figure~\ref{Fig-1}, where we collect some of the main physical parameters of the stars that compose the binary. Among them are the mass and radius determined by \citet{ODonoghue+2003} with values of 0.78 $\pm$ 0.04 M$_{\odot}$ and 0.01 $\pm$ 0.01 R$_{\odot}$ for the WD star, and values 0.43 $\pm$ 0.04 M$_{\odot}$ and 0.42 $\pm$ 0.02 R$_{\odot}$ for the M-class MS dwarf star. On the other hand, we calculated the luminosity of both stars using the temperature provided by \citet{ODonoghue+2003} and their respective radii, finding approximately 0.004 L$_{\odot}$ and 0.01 L$_{\odot}$ for the two stars, respectively.

\citet{Parsons2016} find certain prominences in their spectra that are maintained over time. These arise from the M-dwarf and appear to be locked in stable configurations within the binary system; their manifestations are found to last for more than a year. Besides the WD eclipse, the binary's light curve shows a small reflection effect at blue wavelengths, and ellipsoidal modulation at redder wavelengths \citep{Parsons+2010}.

{Several attempts have been made to elucidate the underlying cause of the substantial and erratic period variations observed in the binary eclipses of QS Vir. The energy available in the secondary star was calculated by \citet{Qian+2010}, who determined that it was insufficient to account for the observed large-amplitude $O-C$ variations\footnote{residuals between the calculated (C) and observed (O) mid-eclipse times.} through Applegate's mechanism. Instead, these authors proposed a combination of a significant continuous reduction in the binary's orbital period and the presence of a circumbinary planet with a mass of approximately 7 M$_\mathrm{Jup}$.}
{However, \citet{Parsons+2010} obtained new eclipse data that suggest a broader and more eccentric orbit for the solitary companion, as documented in Table \ref{Table:aut}.  \citet{Almeida+2011} subsequently presented an alternative data fit, which incorporated two circumbinary objects: a giant planet (approximately 0.009 $M_\odot$) and a brown dwarf (approximately 0.056 $M_\odot$) in orbit around QS Vir. Nevertheless, the significant deviation occurring near cycle number 30\,000 requires at least one unseen massive companion being in a highly eccentric orbit, resulting in an orbital-crossing architecture.}

\citet{Horner+2013} showed that the two circumbinary companions proposed to orbit QS Vir are dynamically unstable on timescales of less than 1000 years. This applies across the entire range of orbital elements that provide a plausible fit to the observational data of \citet{Almeida+2011}. Therefore, the proposed planetary system with two planets orbiting QS Vir is unlikely to exist.

interpretation of the observed ETVs in QS Vir remains remarkably challenging. According to \citet{Bours+2016}, there are 86 published mid-eclipse times, and they added an additional 24 previously unpublished measurements as part of the eclipse timing programme presented in their work. In addition, from the latest eclipse times, the authors report another local or absolute maximum in the $O-C$ residuals, similar to the $O-C$ variations observed around cycle numbers 5000 and 20000.

In the present work, we revisit the orbital fits for QS Vir using an existing dataset from previous research. We examine a total of 105 mid-eclipse times for QS Vir, comprising 86 that have been publicly reported, and 19 additional ones that have not been analyzed before \citep[see their Table A.41]{Bours+2015}.
This dataset provides the cycle number, mid-time transits in Barycentric Julian Date (BJD) with the corresponding errors, and the corresponding reference for the data. In total, the dataset considered here contains 105 observations covering 22.9 $yr$ of observations.
The effect and contribution to the $O-C$ amplitude due to various other effects are hard to estimate, and in the following we assume that the observed $O-C$ variations are due to a single companion.

This paper is structured as follows. In Section~\ref{sec:methods}, we describe the procedure to obtain the named $O-C$ curves from mid-transit times, and the procedure applied to obtain orbital fits. In Section~\ref{sec:errors}, we analyze the errors considering different approaches, paying attention to the robustness of the solutions over the years and discuss the sensitivity of the fits as a function of the number of timing measurements. Finally, in Section~\ref{sec:conclusions}, we present our conclusions.

\section{Methods}\label{sec:methods}

In this section, we present the method used to construct the named $O-C$ signal, which consists in calculating the difference between the observed  mid-eclipsing time with either linear or quadratic time dependence. 

\subsection{Linear ephemeris versus quadratic ephemeris}\label{sec:oc}

An isolated binary star without considering additional effects (such as tides, pericentric movement, and angular momentum exchange) has a constant orbital period. The presence of an additional perturbing body introduces period changes in the binary orbit.

The models that predict the mid-transit times of the observed eclipses $T_{\rm obs}$ are the linear, sinusoidal LTT, and quadratic ephemeris \citep[although, more sophisticated models can be considered according to][]{Hilditch2001}. We consider a combination of these three effects to build the $O-C$ signal, which is the difference between the observed times ($T_{\rm obs}$) and the calculated times ($T_{\rm calc}$):\footnote{A word of caution: In the case of poor sampling of mid-eclipse times, the analysis can result in erroneous results. We here assume that we are working on a sufficiently sampled signal.} 

\begin{align}
(O-C)    &= T_{\text{obs}} - T_{\text{calc}}, \quad \text{with} \nonumber \\
T_{\text{calc}} &= \begin{cases} 
          T_{0} + P_{\text{bin}}l & \text{linear ephemeris} \\
          T_{0} + P_{\text{bin}}l + \beta l^2 & \text{quadratic ephemeris},
      \end{cases}
\end{align}

where $T_0$, $P_{\rm bin}$, and $l$ are the initial epoch (or mid-eclipse at cycle $l$=0), the orbital period of the binary, and the cycle number (l = 0, 1, 2,..). The factor $\beta$ in the quadratic ephemeris {models a secular change in the binary orbital period and can be thought of as a factor that describes the damping (change) of the binary period due to mass transfer, magnetic braking, and gravitational radiation \citep{Gozdziewski+2012}. We used 105 mid-eclipsing times of QS Vir published by \cite{Bours+2015}, because this latter is the most updated publicly available dataset that can constrain the signal. For mid-eclipsing times, we used Barycentric modified Julian Dates, {with Barycentric Dynamical Time}, {BMJD (TDB)}, which according to \citet{Bours+2016} are corrected for the motion of the Sun around the Barycentre of the Solar System\footnote{equivalent to (BJD-2,450,000.0), {but for sake of simplicity we use BMJD.}}. The results of the fits with both models $T_{\rm calc}$ are
\begin{eqnarray}
  T_\text{linear}    = 48689.14152(11) + 0.1507574746(30)  l     \label{eq:fit} \\
  T_\text{quadratic} = 48689.14177(14) + 0.1507574360(13)l  \nonumber \\
                + 7.37(2.47) 10^{-13}  \; l^2 \label{eq:fit2}
.\end{eqnarray}
Depending on which ephemeris is considered, the shape of the $O-C$ curve drastically changes (see different symbols in the top panel of Figure \ref{Fig-3}). 

\begin{figure}[h]
\begin{center}
\includegraphics[width=1\columnwidth ]{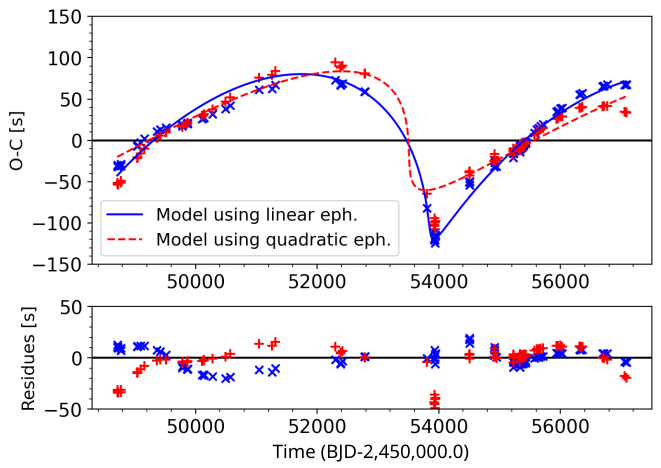}\\
\caption{{Best orbital fits and residuals}. Upper panel: $O-C$ diagram of QS Vir. The crosses and plus symbols represent the $O-C$ values built with the linear and quadratic ephemeris, respectively ($O-C$=$T_{calc}$-Eq(\ref{eq:fit})). The solid line and the dashed line represent the models fitted using the two models as described in the text. Lower panel: Residuals using the same symbol and color coding as the top panel.}
\label{Fig-3}
\end{center}
\end{figure}

\begin{figure}[h]
\begin{center}
\includegraphics[width=0.95\columnwidth]{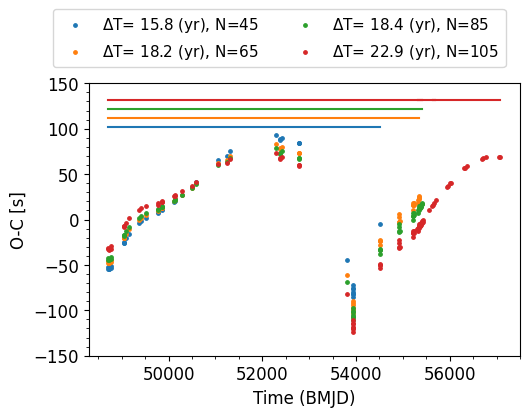}
\includegraphics[width=0.95\columnwidth]{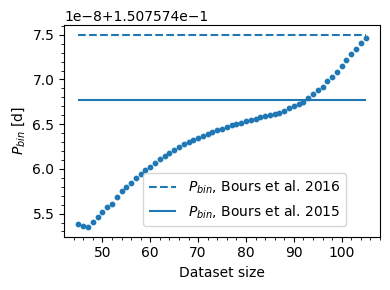}\\
\caption{Dependence of the shape of the $O-C$ and the period of the binary on the dataset considered. Top panel: $O-C$ signal produced using four different datasets that mimic available observational data. Data point colors correspond to different datasets and horizontal lines at the top help to identify the corresponding observational baseline ($\Delta T$). Bottom panel: Determination of the binary period from the linear ephemeris as a function of the dataset. The horizontal solid and dashed lines represent the values of $P_{bin}$ given by \cite{Bours+2015} and \cite{ODonoghue+2003}, respectively. $P_{bin}$ changes depending on the dataset selected.
}\label{Fig-timeinterv}
\end{center}
\end{figure}

\begin{table*}[]
\begin{center}
\caption{Best Keplerian orbital fits for a circumbinary companion around QS Vir.}
\begin{tabular}{lcccccc|c|ccc}
\hline\hline\noalign{\smallskip}
        & $K$       &   $e$     & $\omega$  &   $t_0$   & P     & Z  &  ${\chi^2_\nu}^{1/2}$  &  a & $m_P$ \\
        & [s]    &  & [$^\circ$]        & [BMJD]  &[yr]&        [s] &       &    [au] &   [$M_J$]        \\
Linear & 148.66 & 0.91  & 215.95 & 53849.16 & 16.69 & -99.73
& 14.91 & 7.06 & 57.71 \\


MCMC & $\bullet_{-2.42}^{+26.72}$  &  $\bullet_{-0.01}^{+0.04}$  & $\bullet_{-5.52}^{+0.53}$ & $\bullet_{-4.55}^{+1.63}$ & $\bullet_{-0.03}^{+0.05}$ & $\bullet_{-9.53}^{+1.46}$ &  &  & \\
Grid & $\bullet_{-38.52}^{+145.44}$  &  $\bullet_{-0.17}^{+0.07}$  & $\bullet_{-24.09}^{+9.08}$ & $\bullet_{-21.91}^{+32.26}$ & $\bullet_{-0.42}^{+0.47}$ & $\bullet_{-34.74}^{+39.94}$ &  & &  \\

\hline
Quadratic & 272.20 & 0.97 & 186.18 & 53519.54 & 17.39 & -16.87 & 32.84 & 7.15 & 105.29 \\

\hline
\end{tabular}
\label{Table:lq}
\end{center}
    \tablefoot{Orbital elements are osculating, considering a central star with the total mass of the system, i.e., $m_{\rm bin} [M_{\odot}] = M_0+M_1=0.78+0.43 = 1.21$ \citep{ODonoghue+2003}; the Semi-major axis, $a$, and companion mass, $m_P,$ were calculated assuming the same orbital inclination as the binary, $i$ = $75.5^\circ$ \citep{ODonoghue+2003}. The errors for the quadratic fit are not calculated because the residuals doubled the corresponding ones from linear ephemeris.} 
\end{table*}

\citet{Qian+2010} were the initial proponents of modeling an unseen companion to QS Vir. These authors used a limited dataset consisting of 36 timing measurements collected over a period of 17 years to describe the orbital behavior of QS Vir.
The addition by \citet{Parsons+2010} of 16 new mid-time eclipse observations led to significant changes in the $O-C$ binary period, and in $O-C$ shapes. Consequently, best-fit orbital parameters and masses changed between authors (see our Table  \ref{Table:aut}). More importantly, we note that for each dataset considered, the binary period needs to be redetermined, therefore producing slightly different $O-C$ signals. This could explain the discrepancies observed in the $O-C$ signals published in previous works. To illustrate this, Figure \ref{Fig-timeinterv} presents four $O-C$ signals constructed using the first 45, 65, 85, and 105 mid-time eclipses in chronological order. Each signal is depicted in a distinct color, revealing variations in the shape, maximum, and minimum of the $O-C$. These changes indicate different orbital parameters for each corresponding fit.

\begin{table}[h]
\centering
\caption{Parameters for the planets proposed for QS Vir}
\begin{tabular}{l@{\hskip3pt}c@{\hskip3pt}c@{\hskip3pt}c@{\hskip3pt}c@{\hskip3pt}c@{\hskip3pt}c@{\hskip3pt}c@{\hskip3pt}cc}
\hline\hline\noalign{\smallskip}
Parameter   &   $a$      &  $b$ &  $c$  & d & $\textbf{e}$  \\
\hline
P [yr]           &  7.8  &  14  &  $14.40/16.99$   & $4.78/18.96$ & \textbf{16.69}         \\
e                & 0.37  &  0.9   &  $0.62/0.92$   & $0.10/0.96$ &  \textbf{0.91}         \\
$\omega$[°]      &  38    &  -   & $180/219$       & $337/206$    & \textbf{215.95}         \\
$t_{0}$ [BMJD]  &  48687.5&  -     & -  & -                     & \textbf{53849.16}        \\
$Z$ [s]    & -&  - &  -  & - & \textbf{-99.7}        \\
\hline
M $\sin$ i [M$_J$]    & 6.4 & 54 & 8.04, 52.3 & 6.3, 57.71 & \textbf{57.71}        \\

\hline
\end{tabular}
\label{Table:aut}
Ref.: $^a$ \cite{Qian+2010},  $^b$ \cite{Parsons+2010}, $^c$ \cite{Almeida+2011}, $^d$
\cite{Pereira_2018}, $^e$
\textbf{This work}. 
\label{tabla1}
\end{table}

To test the robustness of the underlying model, one could successively remove the last measured timing and measure the resulting best-fit parameter from linear ephemerids. Furthermore, successive fitting of linear ephemeris shows that the period of the binary changes constantly as a function of data points, as can be seen in the bottom panel of Figure \ref{Fig-timeinterv}. Interestingly, the two values for the binary period widely used for QS Vir are $P_{\rm bin}=0.150757475$ \citep{ODonoghue+2003} and $P_{\rm bin}=0.150757467717$ \citep{Bours+2015}, which are consistent with the dataset presented in this work.

\subsection{Best orbital fits}\label{sec:bestfit}


The presence of a third body produces a signal that can be modeled in the $O-C$ diagram, minimizing the residuals. The signal $\tau$ is obtained from the subtraction of mid-time transits with either the linear or quadratic fit, which can be modeled with a Keplerian orbit. The signal $\tau$ depends on $K$, $P$, $e$, $\omega$, $t_0$, and $Z,$ which represent the semi-amplitude, period, eccentricity, argument of periapsis, time of passage at periapsis, and the origin value for the movement of the baricenter, respectively \cite[we refer to the original formulation of ][for an explanation of the meaning of $Z$]{Woltjer+1922}. Thus, as derived in \citet{Gozdziewski+2012}, $\tau$ can be written as
\begin{equation}
\tau = K \left( \sin \omega \left( \cos E(t) - e \right) +
      \cos \omega \sqrt{1-e^2} \sin E(t)
     \right) + {Z,} \label{eq:tau}
\end{equation}
with
\begin{equation}
     K = \left(\frac{1}{c} \right) \frac{m_P}{m_P+m_{\rm bin}} a \sin i,\end{equation}
where ${m_P}$ is the companion mass, $a$ the companion semi-major axis, $i$ the companion inclination, $E(t)$  the eccentric anomaly, and $m_{\rm bin}$ is the mass of the binary. For a best-fit parameter of $K$, there exists a set of values of $\{m_{P}, a \sin i$, $m_{bin}\}$ that produces the same amplitude of the signal.

To calculate the best-fit parameters, we minimize a certain function of the residuals, which is defined as a statistical measure of the goodness of the fit. Assuming a Gaussian distribution for the errors, the goodness-of-fit statistics usually used is $(\chi^2_\nu)^{1/2}$. This quantity is defined as follows:
\begin{equation}
      (\chi^2_\nu)^{1/2} = \frac{1}{(N-M)} \sum_{i=1}^{N}\frac{ ({O-C}-\tau)^2_{t_{i}}}{\sigma_{t_{i}}^2} \label{eq:chi2}
,\end{equation}
which depends on $(\text{O-C}-\tau)_{t_i}$, $\sigma_{t_i}$, $N=105$, and $M=6$, which are the difference between the observer $O-C$ signal with respect to the modeled $\tau$; the uncertainties for the time of eclipsing $t_i$; the number of observations; and the parameters to be determined, respectively. The quantities $\sigma_{t_i}$ are usually referred to as weights of the observations.

The strategy of fitting the $O-C$ signal using a global exploration of the parameter space with a genetic algorithm followed by a local search for the minimum has already been shown to be successful in modeling the ETV in many works \citep[see e.g.,][]{Beuermann2012, Almeida2013, Almeida2020, Brown-Sevilla2021, Er+2021}. Here, we employ the same strategy, with particular attention to the robustness of the fit.
We used two different and sequential subroutines to calculate the best Keplerian fit. First, we used a genetic algorithm with a population of $150$ members (with random initial conditions in the parameter space), which evolved over approximately $1000$ generations. We explored the hyper-parameter options and found optimal values for a mutation rate dithering based on the generation number of ($0.025, 0.5$), a relative tolerance for convergence of ($0.001$), and a strategy to evolve the population, retaining the best solution among new generations. We ran the algorithm and found the global minimum, although for other systems and/or datasets, these numbers could change depending on the coverage periods of the dataset \citep[see e.g,][for radial velocity applications]{Beauge_2008, Giuppone+2011}. As genetic algorithms are only exploratory tools, they only guarantee a certain proximity to the global minimum of the fitness function, and not a precise value. Finally, we used a simplex subroutine to improve the result \citep[Nelder-Mead algorithm,][]{Press+1992} with a tolerance of $10^{-5}$ to assure the convergence of the goodness-of-fit evaluation. We find a global minimum with parameters, $\mathcal{A}_{BF}$. Our code relies on the algorithms {\tt scipy differential evolution}\footnote{\href{https://docs.scipy.org/doc/scipy/reference/generated/scipy.optimize.differential_evolution.html}{scipy differential evolution}} and {\tt scipy optimize.minimize}\footnote{\href{https://docs.scipy.org/doc/scipy/reference/generated/scipy.optimize.minimize.html}{scipy optimize minimize}} from the {\tt SciPy} Python packages \citep{scipy}. 

Our best-fit values are shown in Table \ref{Table:lq}. With these parameter values, we reconstruct the $\tau$ signal using lines in the top panel of Figure \ref{Fig-3}. The lower panel in the same figure shows the distribution of the residuals. The points obtained with the quadratic ephemeris fit (in red) produce a signal $O-C$ that covers more than one period, although the maximum values ($\sim 100$ s) observed at $T \sim 53000$ are not similar to those observed at $T \sim 57000$ ($\sim 50$ s). This asymmetry prevents us from finding a good solution with a single frequency. Consequently, the values of the goodness-of-fit statistic are downgraded $(\chi^2_\nu)^{1/2}\sim32.84$ (see the values in Table~\ref{Table:lq}). 
The residuals for the linear ephemeris show a sinusoidal behavior with a lower value of $(\chi^2_\nu)^{1/2} (\sim ~14.91)$. This solution corresponds to a mass of the unseen additional body of equal to $m_P=57.71$ $M_{\rm jup}$. 
{We conducted a Lomb-Scargle \citep{Lomb1976,Scargle1982} periodogram analysis to assess the significance of periodic signals. The Lomb-Scargle algorithm is suitable for unevenly spaced data and has been applied in a similar context in the literature \citep{Cortes2020,Burt2016PhDT}. Initially, the Lomb-Scargle analysis was applied to the $O-C$ values derived from the linear ephemeris, revealing a single periodic signal at approximately $16.43$ years with a signal-to-noise ratio ($S/N$) of around $3.5$ — a result consistent with our best-fit period of $16.69$ years. Subsequently, following our best-fit analysis, the Lomb-Scargle periodogram applied to the new $O-C$ data identified two periods with peaks at approximately $4.98$ and $7.34$ years, each associated with a lower $S/N$  of approximately $1.5$. We obtain our results by running the Lomb-Scargle implementation within the {\sc astropy v6.0} package \citep{astropy}}.

For the remainder of this work, we only consider the described $O-C$ signal model for a linear ephemeris. We also explore the solutions given by the quadratic ephemerids but do not show them here as they provide less accurate fits to the observed data.

\subsection{Orbital stability of solutions}
{In order to analyze the stability of the best-fit solutions described in the following sections, we performed long-term $n$-body integrations of each solution. The simulations were performed with a Bulirsch-Stoer integrator with adaptive step size, which was modified to independently monitor the error in each variable; this imposes a relative precision of better than $10^{-12}$. The computations are stopped when the distance from the planet to any star is less than two times the sum of their radii, or when the planet is ejected from the system after scattering between bodies ($>10\,\mathrm{au}$). We use Jacobi elements for our initial conditions. For each simulation, we calculate the averaged MEGNO ($\mathrm{\langle Y \rangle}$) chaos indicator \citep[see][for details]{Cincotta.Simo.1999}. As the value of this indicator may be sensitive to the integration time span \citep[][]{Cincotta2003, Hinse2010}, we verified its value for $10^3$ and $10^4$ years. We study only coplanar orbits, and initial angular orbital elements for the binary were set equal to zero. We have previously
examined  stability and chaos for planets around binaries \citep{Gianuzzi+2023A&A}. However, it is worth noting that the present work is considered an extreme case due to the high eccentricity of the third body $(e > 0.8)$.}

\section{Strategy to determine errors in orbital fits}\label{sec:errors}

\begin{figure*}
\begin{center}
\includegraphics[width=0.9\textwidth]{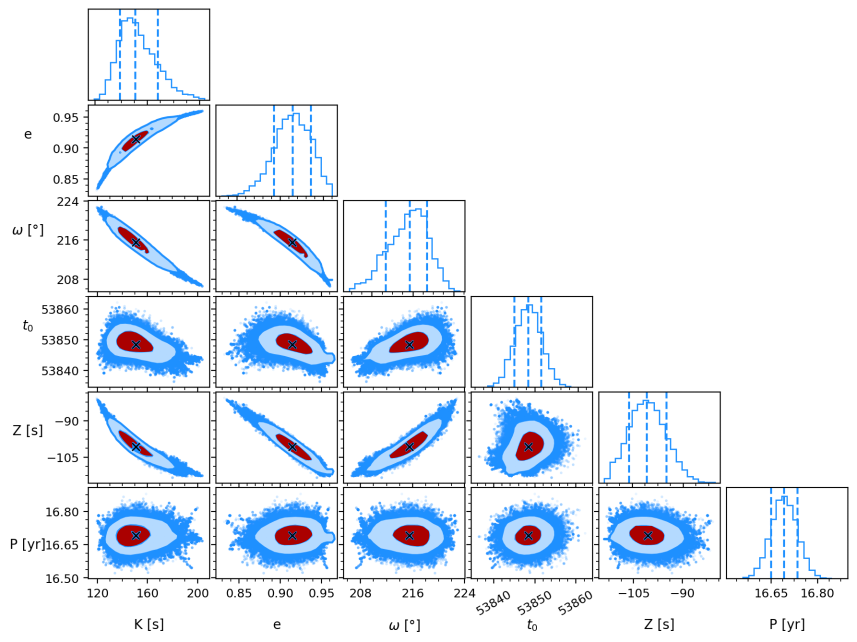}
\caption{Results of MCMC chains selecting pairs of parameters. We estimate the uncertainties using the posterior distribution of MCMC chains. Mean values, i.e., best-fit solutions, are plotted with black crosses. Level curves indicate $1\sigma$ (dark red) and $2\sigma$ (light blue) levels of posterior distributions. We note that for 2D histograms, these values correspond to roughly 39\% and 86\% confidence levels of the mean. 
In the top panels, we show the posterior distributions with their quantile levels at 0.14, 0.5, and 0.84. Here, $t_0$ is given as a BMJD.} \label{Fig-MCMC}
\end{center}
\end{figure*}

In the following, we briefly outline and follow two methods for the computation of parameter uncertainties. This approach {provides} a critical evaluation of the robustness of the uncertainties. If the same confidence interval is reproduced from two different methods, we lend trust to the derived uncertainties. Otherwise, we consider that there is little or no consensus at all between the two methods, and the more conservative confidence interval is then adopted. Namely, we use the Markov chain Monte Carlo (MCMC) algorithm and other proven alternative strategies that help to understand the errors in this kind of minimization \citep{Beauge_2008}.


\subsection{Parameter uncertainties from MCMC}

MCMC methods are designed to sample the errors from the posterior probability density function (PDF) by utilizing the likelihood function, even in parameter spaces with high dimensionality \citep[see][for in-deep discussion]{Foreman+2013}. In simpler terms, this method estimates errors by sampling regions near the best-fit solution using the posterior distribution. To determine uncertainties, we obtain the $1\sigma$ interval of the 1D projections of the sampling onto the parameter subspace ($\mathcal{A}$=$K,e,\omega,Z,t_0,P$). We used the \textit{emcee} implementation described in \cite{Foreman+2013}, configuring the number of MCMC steps as $N_{MCMC}$, the number of chains (or \textit{walkers}) as $\mathcal{W}$, and the initial distribution of walkers, \textit{$\beta_{MCMC}$}. The best-fit parameters obtained from the linear fit model (see Table \ref{Table:lq}) served as priors in the MCMC.

Following the recommendations of \cite{Foreman+2013}, we employed an ensemble of 64 walkers ($\mathcal{W}=32 - 64)$) and ran the chain for $N_{MCMC}=100\,000$ steps. The initial distribution of walkers was set as $\beta_{MCMC}=0.01$ (except for $t_0$, where $\beta=0.0001$). We discarded the first 7\,000 steps as part of the burn-in stage and estimated uncertainties using the posterior distribution. 
The dispersion point panels shown in Fig \ref{Fig-MCMC} depict parameter correlations, while the histograms represent posterior probability density functions for the selected parameters. The uncertainties are calculated as the range of values encompassing 66$\%$ of the mean (indicated on each histogram). This figure offers valuable insights into parameter space exploration, facilitating a comprehensive understanding of uncertainties and correlations among model parameters. Notably, strong correlations are observed between ($\omega,K$), ($e,K$), ($e,\omega$), ($Z,K$), ($Z,e$), and ($Z,\omega$). Visualizing these pairwise relationships helps in identifying optimal parameter combinations, thereby enhancing the robustness of our analysis. Despite the absence of multimodal posteriors, the $O-C$ signal models exhibit significant parameter correlations, making the application of the MCMC method to explore the entire parameter space particularly challenging \citep{Foreman+2013, Marsh+2014}.

We conducted a thorough evaluation of the MCMC analysis to ensure convergence and reliability of the results. This involved carefully examining convergence diagnostics, increasing the number of iterations, and adjusting the burn-in period to ensure it was of sufficient length. Despite these efforts, the obtained results remained unchanged. It is important to note that the accuracy of uncertainty estimates from MCMC chains relies on the assumption that the model being fitted to the data accurately captures the underlying complexities of the observed phenomena. 

We note that MCMC is the most commonly method adopted by observers when reporting orbital fits for regularly sampled data. Therefore, as these errors appear to be underestimated according to the orbital fits reported by different authors, in the following we present a complementary approach that allows us to better estimate these errors.

\subsection{Stability of MCMC solutions}

To analyze the stability of the results of MCMC chains, we
first calculated the Mahalanobis distance $r$ of each solution. This quantity represents the distance between each point $\mathbf{x}$ and the mean $\mathbf{\mu}$, and is defined as
\begin{equation}
    r(\mathbf{x}):= \sqrt{\mathbf{x}^\mathrm{T}\mathbf{\Sigma}^{-1}\mathbf{x}},
\end{equation}
where $\mathbf{\Sigma}$ is the symmetric covariance matrix. In the 1D case, this quantity reduces to the absolute value of the standard score ($\sigma$), and the distribution reduces to a univariate normal distribution.

We generated a random sample of 1000 solutions out of the $r$-closest 67200 solutions (30\% of the total), and integrated them for {5000}$\,$years. Figure~\ref{MCM_megno} shows the averaged MEGNO value $(\langle\mathrm{Y}\rangle)$ of each integrated solution as a function of the Mahalanobis distance. All solutions have $ | \langle\mathrm{Y}\rangle - 2.0 | < 0.005$, which means that all the orbits are regular. We do not find a correlation between $\langle\mathrm{Y}\rangle$ and $r$.

\begin{figure}[h]
\begin{center}
\includegraphics[width=0.9\columnwidth]{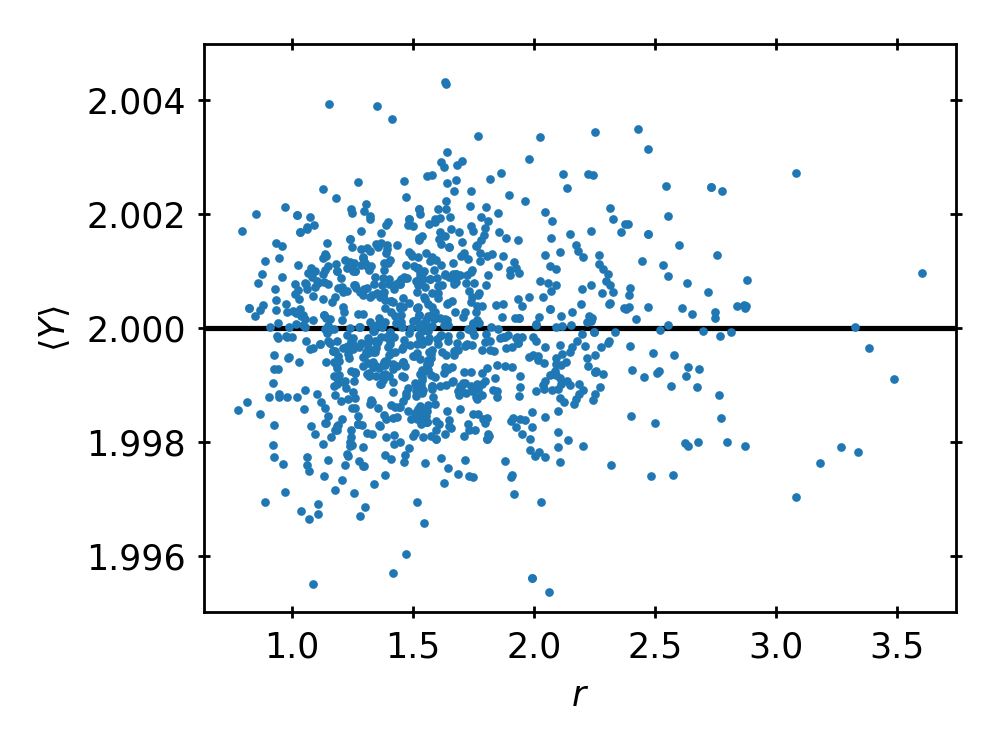}
\caption{Averaged MEGNO as a function of  Mahalanobis distance. The horizontal line denotes $\langle\mathrm{Y}\rangle=2$.
}\label{MCM_megno}
\end{center}
\end{figure}

\begin{figure*}[h]
\begin{center}
\includegraphics[width=0.9\textwidth]{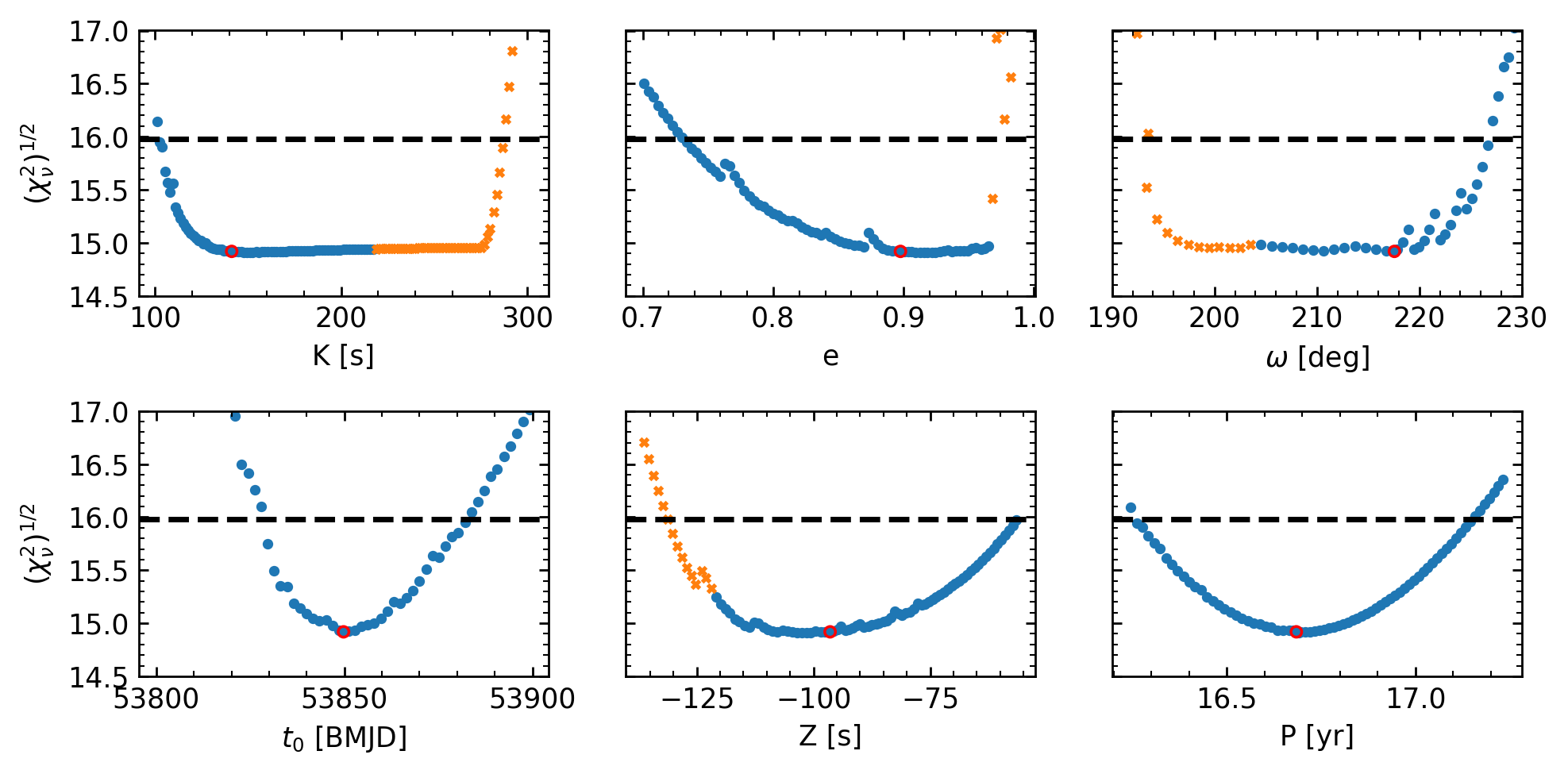}
\caption{Goodness-of-fit $(\chi^2_\nu)^{1/2}$ as a function of a fixed parameter $\mathcal{A}_i$, while letting the remaining parameters vary freely. The red circle marks the best-fit value ($\mathcal{A}_{BF}$ as determined from the genetic-simplex method). The dashed line marks the value of $(\chi^2_\nu)^{1/2}_{1\sigma}$ defined in Eq. \ref{eq:chi2sigma}. The intersection of $(\chi^2_\nu)^{1/2}_{1\sigma}$ with the blue points defines the ${1\sigma}$ confidence interval for the parameter $\mathcal{A}_i$ (see the results in Table \ref{Table:lq}).
Orange circles correspond to chaotic orbits, with $\langle\mathrm{Y}\rangle \gtrsim 2.1$ (see the text).
}\label{Fig-grid}
\end{center}
\end{figure*}

\subsection{Parameter uncertainties from a grid search around the best fit}\label{sec:3.2}
Starting from the best-fit solution $\mathcal{A}_{BF}$, we selected each parameter individually and evaluated the change in $(\chi^2_\nu)^{1/2}$, calculating a new $(\chi^2_\nu)^{1/2}$ , where we only fixed the new value of a selected parameter, leaving the others the possibility to change. For example, in top left panel of Figure \ref{Fig-grid}, we show subsequent fits from the best-fit solution (red cross) in the amplitude $K$. {Every new $K_i$ is calculated} in a grid of values of $K_i=K+\delta,K+2\delta,..,K+n\delta$, where $\delta$ is a small parameter. We repeated the procedure {for positive and negative values of $\delta$,} and plot the values of $(\chi^2_\nu)^{1/2}$ with respect to $K_i$. We note that the remaining parameters from $\mathcal{A}$ are fitted and this is a projection in the plane $K-(\chi^2_\nu)^{1/2}$. We observe a depth minimum starting at $(\chi^2_\nu)^{1/2}=14.91$ and increasing as we depart from the best fit. We repeated the procedure for the remaining five parameters of $\mathcal{A}$ and plot the results in the panels of Figure \ref{Fig-grid}. 

To estimate the uncertainties, we employed a method based on the $\chi^2_\nu$ statistic and calculated the $1\sigma$ confidence levels. Due to the nonlinearity of the fitness function, conventional confidence ellipsoids cannot be directly applied in this case. Following the approach described in \cite{Beauge_2008}, we set $\nu = N - M$, where $N$ is the number of data points and $M$ is the number of model parameters. By determining the mean ($\nu$) and variance ($2\nu$) of $\chi^2_\nu$, we can approximate the $1\sigma$ confidence level value using the formula:
\begin{equation}
(\chi^2_\nu)^{1/2}_{1\sigma} \simeq (\chi^2_\nu)^{1/2}_{\mathcal{BF}} \left( 1 + \sqrt{\frac{1}{2\nu}} \right) \label{eq:chi2sigma}
.\end{equation}

We reiterate that our $(\chi^2_\nu)^{1/2}_{\mathcal{BF}} =14.91$ (see Table \ref{Table:lq}). To obtain the $1\sigma$ value for each parameter $\mathcal{A}_i$, we perform a numerical analysis to identify the intersection of $(\chi^2_\nu)^{1/2}$ with the $(\chi^2_\nu)^{1/2}_{1\sigma}$ ($\sim$ 15.98)  
at each panel of the $\mathcal{A}_i$ best fits. The resulting estimation of the uncertainties in $\mathcal{A}_i$ is given in Table \ref{Table:lq}. This approach allows a quantitative assessment of the range of $\mathcal{A}_i$ values consistent with the dataset and provides insight into the reliability and confidence associated with the parameter estimation.

Additionally, we show the results obtained from the dynamical study of our solutions in Fig. \ref{Fig-grid}. We identify regular and chaotic orbits with different symbols. However, chaotic and regular orbits are stable in long-term integrations for $1 \times 10^6\,\mathrm{yr}$ (i.e., $\sim 6  \times 10^5$ periods of the unseen companion). The orbits neither escape from the system nor have a close encounter between involved bodies. Remarkably, we observe that the systems with the highest values of $K$ or eccentricity (or lowest $w$ or $Z$ values) have chaotic orbits.

Combining the period of the unseen companion and $K,$ we obtain the mass of the companion (see Eq.~\ref{eq:k}). At this stage, it is relevant to evaluate the stability of these solutions. Figure~\ref{m_megno} shows $\langle\mathrm{Y}\rangle$ as a function of the companion mass for each best fit obtained by varying $K$ in Fig.~\ref{Fig-grid} (top-left panel). Despite a drop at $m_P\sim80\,\mathrm{M}_{J}$, we generally see that larger $m_P$ values result in larger $\langle\mathrm{Y}\rangle$ values. We define $\langle\mathrm{Y}\rangle_c=2.1$ as the averaged MEGNO cutoff value to differentiate chaotic from regular orbits during our integration of 50000 yr. Given this value, most of the solutions with $\langle\mathrm{Y}\rangle>\langle\mathrm{Y}\rangle_c$ (orange crosses in Figs.~\ref{Fig-grid} and \ref{m_megno}) have $m_P>90\,\mathrm{M}_{J}$.

\begin{figure}[h]
\begin{center}
\includegraphics[width=0.9\columnwidth]{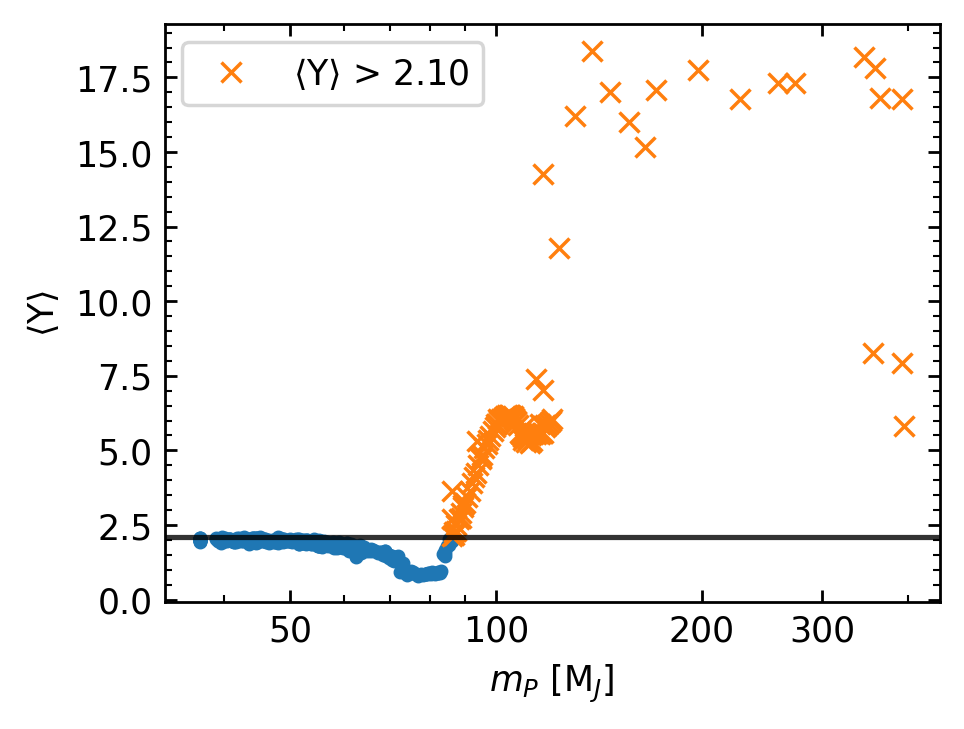}
\caption{Averaged MEGNO as a function of the planet mass in logarithmic scale for integrations over $5 \times 10^{4}$ yr. The horizontal line denotes the critical value $\langle\mathrm{Y}\rangle_c=2.1$, and orange crosses represent integrations with $\langle\mathrm{Y}\rangle>\langle\mathrm{Y}\rangle_c$.
}\label{m_megno}
\end{center}
\end{figure}

In Figure \ref{fig:compare}, we show the long-term evolution of two example orbits with MEGNO values of $\langle\mathrm{Y}\rangle \sim {2.00}$ and $\langle\mathrm{Y}\rangle \sim {17.31}$ at 50000 yr. The initial conditions of both orbits are $a=(7.0758786,\,7.0008565)\,$au, $e=(0.93139948,\,0.99)$, and $m_p=(63.8105,\,273.2768)\,$ M$_J$  for the regular and chaotic orbits, respectively. We choose to show the evolution of the semi-major axis of the best-fit solution and MEGNO in the panels. The orange orbit, although flagged as stable, shows recurrent and irregular variations in semi-major axis (also in eccentricity, but not shown here). In the presence of an additional body, this orbit could lead to a dynamically unstable system.

\begin{figure}[h]
\centering
\includegraphics[width=0.9\columnwidth, clip]{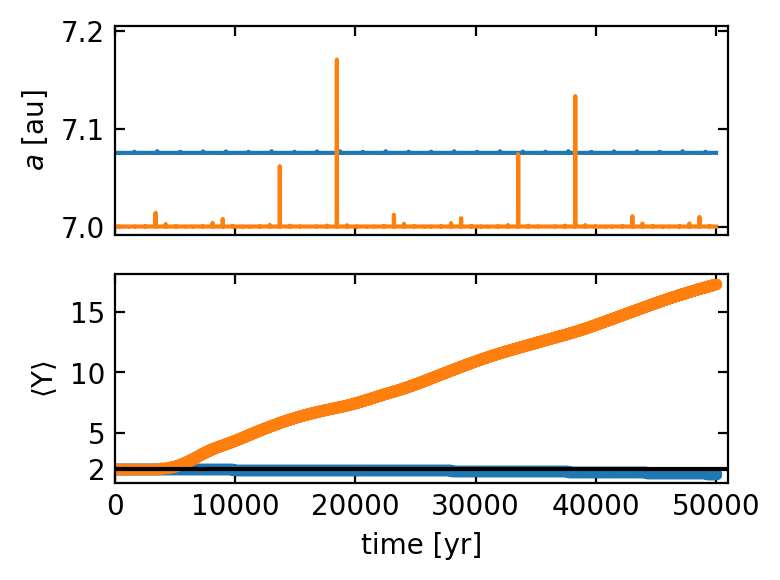}
\caption{Semi-major axis (top) and averaged MEGNO (bottom) as a function of time for a chaotic (orange) and a nonchaotic (blue) system.}\label{fig:compare}
\end{figure}

\subsection{Dependence of the best fit on the choice of observational dataset}

\begin{figure*}[h!]
\begin{center}
\includegraphics[width=0.9\textwidth]{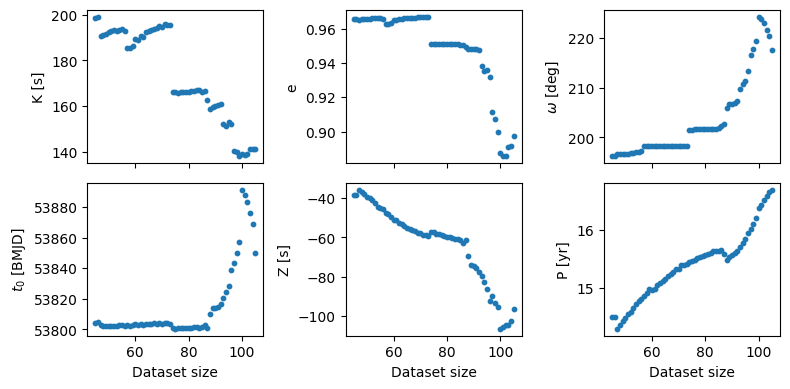}
\caption{Evolution of the parameters fitted as a function of the number of data points used. In the different panels, the value of the best-fit parameter for the third body is shown.}\label{Fig-Ndata}
\end{center}
\end{figure*}
{In this section, we lower the number of observations in the dataset and analyze the resulting variations in the orbital fits. We progressively consider different orbital fits obtained from a reduced dataset where the last point has been removed.} In other words, we first perform the linear fit to obtain the $O-C$ and the  best fit on the full dataset with 105 data points. We then carry out different linear fits to get the $O-C$ with only 104 points (removing the last one, chronologically speaking) and we repeat this procedure until we are left with around 50 points.

\begin{figure}
\begin{center}
\includegraphics[width=0.9\columnwidth]{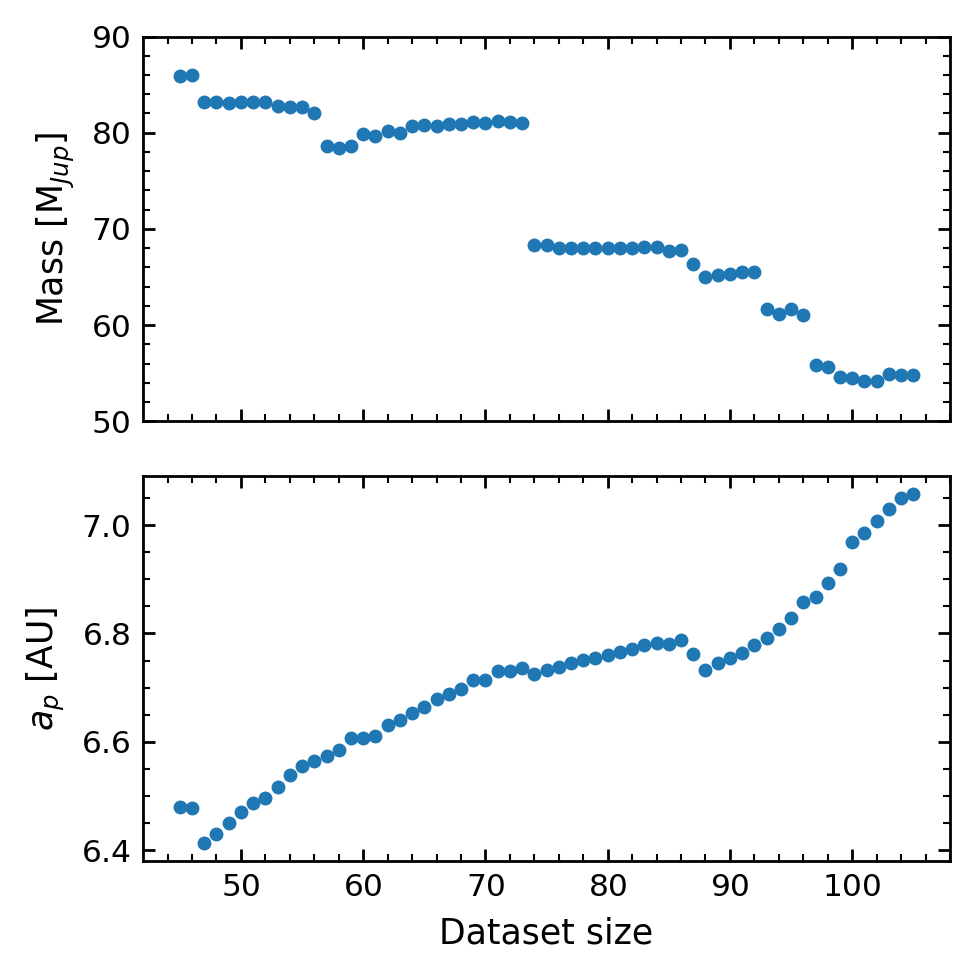}\\
\caption{Mass (top) and semi-major axis (bottom) of the planet as functions dataset size.
}\label{fig:m_and_a}
\end{center}
\end{figure}

We can plot the variation of the parameters $\mathcal{A}$ of each orbit fit as a function of the observation time interval (or the number of data points). If the current solution is robust, we should expect only small and smooth changes in the parameters as a function of the dataset considered. If the solutions do not change, we can expect that the future addition of observations will not significantly change our knowledge of the system. Again, we refer to \citet{Beauge_2008} for a discussion of this method and its application to {the} radial velocity technique.
The results for Keplerian fits are shown in Figure \ref{Fig-Ndata}, where the central eclipse values are ordered chronologically.  We recall that for every new dataset, we need to recalculate the new linear ephemeris (a new $P_{\rm bin}$ value is determined), mimicking the dataset for a given epoch. The best orbital fit was calculated using the simplex algorithm with the starting point using the previous fit in order to speed up the minimization algorithm without considering a global exploration of space parameter.

\begin{figure}
\begin{center}
\includegraphics[width=0.9\columnwidth]{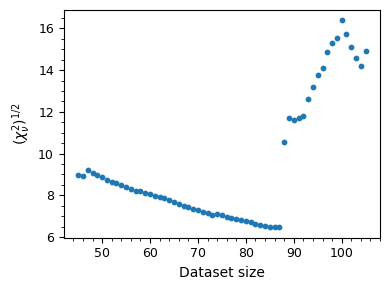}\\
\caption{Dependence of $(\chi^2_\nu)^{1/2}$ on the choice of dataset. $(\chi^2_\nu)^{1/2}$ calculated using the different datasets, with solutions given in Fig. \ref{Fig-Ndata}.}\label{Fig-datasets}
\end{center}
\end{figure}

We note an abrupt change in some parameters when fewer data points are eliminated. For instance, the semi-amplitude of the signal $K$ suddenly increases from $~140 \;s$ to $150 \;s$ at around $\sim 96$ observations. This jump is also observed in the panels of Fig. 9 showing $e$, $Z$, and $\omega$. The argument of periapsis $\omega$ changes by only about$~10\%$, while eccentricity is higher as we consider less mid-time transits. This could explain the higher $e$ values reported in the solutions by former authors. We reiterate that each dataset has its own binary period determined, and that the shape of the $O-C$ curve to be adjusted also changes (see Fig. \ref{Fig-timeinterv}).

Figure \ref{fig:m_and_a} shows the companion's mass and semi-major axis derived from the fitted parameters  as a function of dataset size. We observe a decrease in mass, from $\sim 90\,\mathrm{M}_{\rm Jup}$ to $\sim 50\,\mathrm{M}_{\rm Jup}$, with increasing dataset size. In addition, there is an almost linear increase in the companion's semi-major axis from $\sim 6.4\,\mathrm{au}$ to $\sim 7.05$ au. The binary semi-major axis remains approximately constant at $a\sim5.907279\times10^{-3}\,\mathrm{au}$, with small increments of the order of $10^{-10}$ au.

Successive fitting gives different values of $(\chi^2_\nu)^{1/2}$, as shown in Fig.~\ref{Fig-datasets}. When the dataset size is around 88 observations, there is a sudden increase in $(\chi^2_\nu)^{1/2}$. 
For smaller datasets, there is a smooth change that leads to a decrease in the goodness-of-fit as we consider additional points, starting from $(\chi^2_\nu)^{1/2} \sim 9$ for 50 points until reaching $(\chi^2_\nu)^{1/2} \sim 6$ for 88 points (see the denominator in Eq.~\ref{eq:chi2} to understand the behavior of $(\chi^2_\nu)^{1/2}$ as we increase $N$). For datasets with more than 88 points, the $(\chi^2_\nu)^{1/2}$ increases drastically and presents values around $14$. Therefore, we conclude {for the most recent} mid-time observations that solutions are scarcely distributed and are unable to restrict the possible orbits to a unique solution. In order to comprehend the discrepancy observed in $(\chi^2_\nu)^{1/2}$ for a dataset size of approximately 88, Fig.~\ref{Fig-weights} illustrates the uncertainties associated with each observation. Notably, we observe that the uncertainties in mid-time eclipse times, denoted $\sigma_i$, are potentially underestimated for mid-time transits, particularly when $N > 88$. This discrepancy leads to greater weight being assigned to the new data and has the potential to alter the absolute minimum of orbital fits.

In conclusion, using the data from \citet{Parsons+2010} (58 data points, covering approximately 17.85 years of observations), we note that the orbital fit for determining the hidden companion of QS-Virginis yields completely different results compared to those obtained using 105 data points (spanning nearly 22.9 years). Our findings support the idea that poorly sampled eclipse times, insufficient sampling over long time periods, and/or incomplete coverage of the unseen companion's orbital period can lead to incorrect characterization of the third body. In the specific case of QS-Virginis, acquiring additional data would be desirable in order to establish orbital parameters with greater confidence.


\begin{figure}
\begin{center}
\includegraphics[width=0.9\columnwidth]{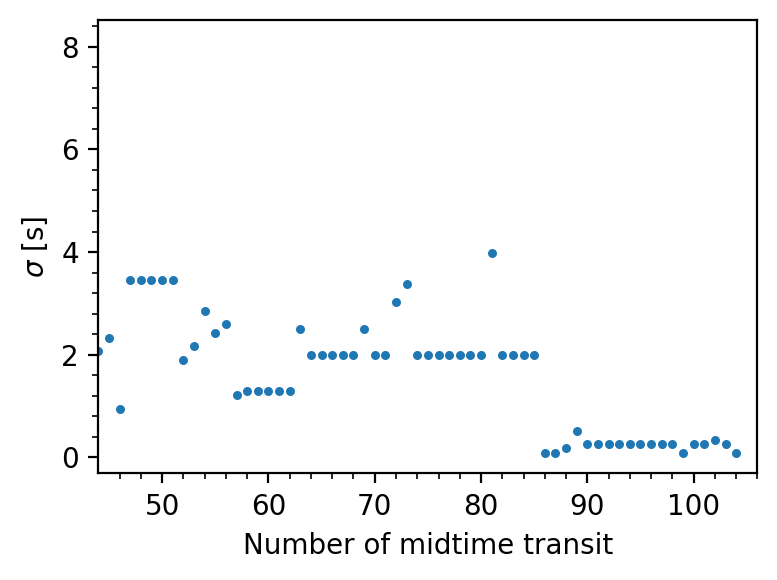}
\caption{Representation of error weights for each mid-time transit as collected by \citet{Bours+2015}. }\label{Fig-weights}
\end{center}
\end{figure}


We integrated the solutions shown in Fig. \ref{Fig-Ndata} performing N-body simulations. All the solutions considered are stable over $10^{4}$ yr and exhibit  
$  log| \langle\mathrm{Y}\rangle - 2.0 | < -1.8$, implying regular orbits. Furthermore, our findings are validated by extensive long-term simulations, where each initial condition was tracked over a span of $10^6$ years ($\sim 2 \times 10^9\,P_{bin} \sim 6 \times 10^4\,P_{pl}$).

Figure $\ref{fig:BFS-Deltas}$ shows the total amplitude of the evolution of orbital elements during the integration, named the $\Delta a=(a_{\max} - a_{\min})$ and $\Delta e=(e_{\max} - e_{\min})$ indicators, for the binary orbit and the planet orbit, respectively. The low values obtained indicate stability for these orbits. We generally observe that $\Delta a$ and $\Delta e$ values are the lowest for the largest dataset. In particular, the results obtained for the fits using all the data points (105) appear to have converged to a steady value. Furthermore, we note that simulations with a dataset of larger than 73 (discontinuity point) exhibit milder variations in orbital elements, suggesting more regular orbits (compared to simulations with smaller datasets).

\begin{figure}
\begin{center}
\includegraphics[width=1\columnwidth]{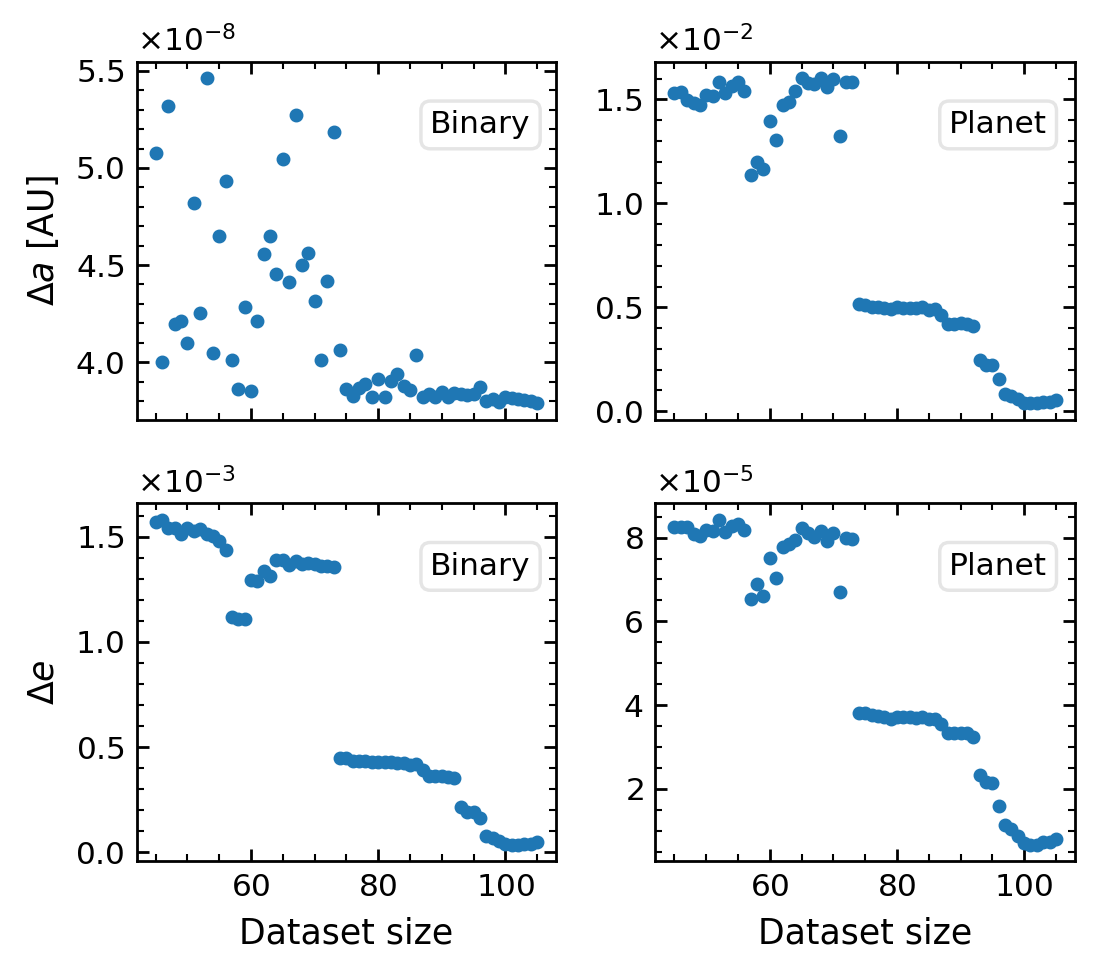}
\caption{$\Delta a$ (top) and $\Delta e$ (bottom) stability indicators of the binary (left) and the modeled companion (right panels) integrated orbits, as a function of dataset size.}\label{fig:BFS-Deltas}
\end{center}
\end{figure}

For each integration, a rough estimation of the lowest pericenter distance ($q_{pl}$) of the planet orbit in a Jacobi reference frame can be computed as:
\begin{equation}
    q_{pl} = a_{min} (1-e_{max}). \label{eq:peric}
\end{equation}
Given that $\delta_a$ and $\delta_e$ of the binary are almost constant, we can estimate the companion's pericentric distance for each simulation when both $\Delta a$  and $\Delta e$ are maximum as
\begin{equation}
    q_{pl} = (a-\Delta a)\bigg(1 - (e + \Delta e)\bigg)
.\end{equation}
In Figure $\ref{fig:BFS-Peric}$, we observe that for dataset sizes $\in [45-73],$ we obtain a nearly constant value of $q_{pl}\sim 49\,\mathrm{R}_\odot$. This is also the lowest value for all the simulations. For dataset sizes of $\in [74-92],$ we reach another nearly constant value of $q_{pl}\sim 73\,\mathrm{R}_\odot$. For dataset sizes $\geq 92$, $q_{pl}$ is always greater than $80\,\mathrm{R}_\odot$, which avoids close encounters with the binary.

\begin{figure}
\begin{center}
\includegraphics[width=0.9\columnwidth]{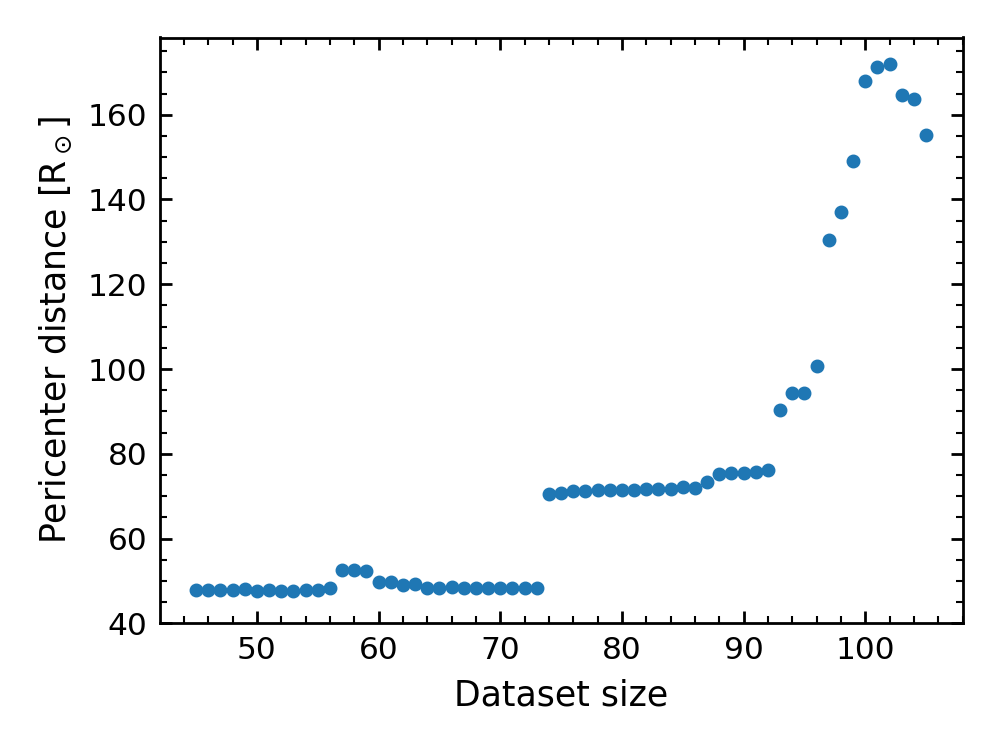}
\caption{Estimated minimum planet pericenter distance as a function of dataset size.}\label{fig:BFS-Peric}
\end{center}
\end{figure}

\section{Discussion and conclusions}\label{sec:conclusions}

We explored different strategies to model the $O-C$ signal present in some of the evolved binary stars. To do so, we designed an algorithm that produces the $O-C$ signals starting from the linear or quadratic ephemeris. We also built an algorithm that models the $O-C$ diagram and obtains the best-fit Keplerian parameters of a third body orbiting the binary. 

We tested our methods on QS Vir because of the intriguing dispersion of best-fit values found in the literature, as well as the availability of an extended baseline of observations spanning approximately 22.9 years. In the case of QS Vir, we reach the following conclusions:

\begin{itemize}
    \item The residuals from a linear ephemeris provide a more suitable model for capturing the $O-C$ signal, while a quadratic ephemeris results in a modulated signal that cannot be adequately modeled with an isolated body around the binary system.

    \item The binary period is found to be modified as a function of the observational data. These modified values roughly agree with those reported by previous authors (see Fig. \ref{Fig-timeinterv}, bottom panel).

    \item The best-fit values for the unseen companion indicate the presence of a low-mass stellar companion (57.71 $M_{\rm jup}$) on an eccentric orbit ($e=0.91^{+0.07}_{-0.17}$), with a range of compatible periods ($P = 16.69 ^{+0.47}_{-0.42}$ years).
    
    \item Parameters such as the companion's period and its argument of periapsis ($\omega$) exhibit significant variations depending on the  size of the observational dataset used.

\end{itemize}

It is important to note that the best-fit values obtained are not statistically robust without a proper error analysis. We successfully tested different strategies for determining the uncertainties of best-fit parameters, and evaluated their variations as a function of the observational data (this allowed us to mimic the observational dataset according to the publication epoch).  In our analysis, we took great care to ensure the suitability of the model and the fidelity of the MCMC methodology in capturing the relevant data features. We thoroughly evaluated the performance of the model and scrutinized the MCMC convergence to ensure both robustness and reliability. The parameter space in these optimization problems is usually a forest of many local minima, and once a solution is found the uncertainties with MCMC are calculated within that particular narrow pit. This is not a failure of the algorithm, but is connected to the intrinsic nature of this problem. 

To further strengthen our error analysis, we introduced an alternative approach known as grid search. This method allows us to systematically search for errors in each parameter $\mathcal{A}i$ that exhibits statistical significance. Specifically, we identified parameters for which their best fits gave $(\chi^2_\nu)^{1/2}$ values lower than the $1\sigma$ value of $(\chi^2_\nu)^{1/2}$. By identifying the interval of possible values of $\mathcal{A}i$, we provide additional insights into the reliability of the parameter estimates.
Our dynamical study of best-fit solutions does not show unstable orbits. However, we find some regions of chaotic solutions, which could shrink the possible solutions of an additional body in the system. However, our study indicates that the current available observations are insufficient to confidently confirm or rule out the presence of  such an additional body.

At face value, and based on the available data, we cannot exclude the presence of a hypothetical fourth body in the outer regions of the QS Vir system. It is worth highlighting that the methodology proposed here is robust enough to tackle the analysis of challenging ETV signals in a more general context. For the enigmatic case of QS Vir, future high-precision measurements will be crucial to reveal its elusive orbital architecture.

\begin{acknowledgements}
We thank the anonymous reviewer who, through a rigorous examination of the manuscript, suggested substantial improvements to this final version.
T.C.H. acknowledges Dr. Jennifer Burt for valuable discussion on the computation of Lomb-Scargle power spectra within PYTHON.
N-Body computations were performed at Clemente Cluster from IATE, Argentina and at the Mulatona Cluster from the CCAD-UNC, which is part of SNCAD-MinCyT, Argentina.
This project has received funding from the European Research Council (ERC) under the European Union Horizon Europe programme (grant agreement No. 101042275, project Stellar-MADE).
\end{acknowledgements}


\bibliographystyle{aa}

\bibliography{bibliography}


\end{document}